\documentclass[showpacs,twocolumn,aps,prb]{revtex4-1}
\usepackage{amssymb}
\usepackage{amsmath}
\usepackage{graphicx}
\usepackage{subfigure}
\usepackage{color}
\usepackage{ulem}

\begin{document}
\title{Intrinsic nonlinear valley Nernst effect in the strained bilayer graphene}
     \author{Ying-Li Wu$^{1}$}
    \author{Jia-Liang Wan$^{1}$}
	\author{Xiao-Qin Yu$^{1}$}
	\email{yuxiaoqin@hnu.edu.cn}
	\affiliation{$^{1}$ School of Physics and Electronics, Hunan University, Changsha 410082, China}

\begin{abstract}
We theoretically analyze the nonlinear valley Nernst effect (NVNE) as the second-order response of temperature gradient through the semiclassical framework of electron dynamics. Our study shows that an intrinsic nonlinear pure valley current can be generated vertically to the applied temperature in the materials with both inversion and time-reversal symmetries. This intrinsic NVNE has a quantum origin from the quantum metric and shows independence from the relaxation time. We find  that the local largest symmetry near the valleys for the nonvanishing intrinsic NVNE is a single mirror symmetry in two-dimensional systems. We theoretically investigate the intrinsic NVNE in the uniaxially strained gapless bilayer graphene and find the intrinsic NVNE can emerge when applying the temperature gradient vertically to the direction
of strain. Interestingly, a transition from the compressive strain to the tensile one results in the sign reversal of the intrinsic NVNE.
\end{abstract}

\pacs{}
\maketitle
\section{Introduction}
The generation of valley current is a vital issue in valleytronics\cite{D.Xiao1,Akhmerov,J. R. Schaibley,Tong. WY,Peng  Rui} in which information is stored and carried through the valley (an extra degree of freedom of electron), rather than electron, in the two-dimensional (2D) crystal with a honeycomb lattice structure. A conventional method to generate valley current is via the valley Hall effect (VHE)\cite{Tong. WY,Peng  Rui,D.Xiao2012}, which describes carriers in two inequivalent valleys ($\mathrm{K}$ and $\mathrm{K}^{\prime}$) moving towards opposite edges perpendicular to an applied electric field due to the nonzero Berry curvature (BC) of the energy band. VHE exhibits a linear dependence on the electric field.

Recently, a subfield research field to valleytronics has emerged, valley caloritronics\cite{Xiaobin Chen,Kapri. Priyadarshini,Zhai. Xuechao}, which explores thermal methods to generate valley current rather than through the electrical approach. Both valley Seebeck effect (VSE)\cite{Zhai. Xuechao,Z. Yu} and valley Nernst effect (VNE)\cite{X.-Q. Yu,Dau. Minh Tuan} have been proposed to thermally generate the valley current.  
Unlike VSE (a longitudinal effect), VNE refers to a transverse phenomenon, in which a linear temperature-gradient-dependent valley current is generated vertically to an applied temperature gradient. Yu \textit{et al.}\cite{X.-Q. Yu} have shown that VNE is also attributed to the BC of the occupied energy band like VHE. However, one might notice that both VHE and VNE originating from the BC as a first-order response to driving forces (electric field or temperature gradient) disappear in the nonmagnetic centrosymmetric materials since the BC becomes zero throughout the whole Brillouin zone when systems are both time-reversal ($\mathcal{T}$) and inversion ($\mathcal{P}$) symmetric. Obviously, the suppression of these linear effects (VHE and VNE) in nonmagnetic centrosymmetric materials greatly limits the selection range of valley materials.

The discovery of the nonlinear anomalous Hall effect (NAHE)\cite{I. Sodemann,Low-2015,Z.Z.Du-2018} as a second-harmonic response to an ac electric field by Sodemann in 2015, which is attributed to BC dipole instead of BC, has attracted broad interests in the exploration of the higher-order transport phenomena stemming from the other quantum geometric quantities beyond BC. A series of novel nonlinear transport phenomena are identified, including nonreciprocal magnetoresistance (NMTR)\cite{D. Kaplan}, intrinsic planar Hall effect(IPHE)\cite{V. A. Zyuzin,Hui Wang}, nonlinear Nernst effect\cite{X.-Q. Yu2,C. Zeng2,Y. Gao1,Harsh Varshney,Liu-2025}, and  nonlinear thermal Hall effect\cite{C. Zeng,D.-K. Zhou,H. Varshney1,H. Varshney2,Zhang-2025}. The quantum metric\cite{D. Kaplan,Hui Wang,Wang1,Zhuang,N.-Z. Wang,A. Gao}, a quantum geometric quantity describing the relative distance between quantum states, shows to play a significant role in NAHE, NMTR and IPHE.

Unlike the BC, the quantum metric can exist in the systems with both $\mathcal{P}$- and $\mathcal{T}$- symmetries\cite{K. Das}. 
Rooting in the quantum metric, Kamal Das \textit{et al}\cite{Das Kamal}. have recently proposed a new type of valley effect, namely nonlinear valley Hall effect (NVHE), to generate valley current flowing vertically to the electric field in tilted Dirac materials containing both $\mathcal{T}$ and $\mathcal{P}$ symmetries. They show that NVHE emerges from an electric-field correction to the BC as a second-order response to the electric field and is independent of the relaxation time $\tau$. Subsequently, NVHE  has also been predicted in a bilayer transition metal dichalcogenide with $\mathcal{P}$ symmetry\cite{Z.-C. Zhou}.  These two works show that a pure valley current can also be generated perpendicularly to the applied electric field
in the nonmagnetic centrosymmetric materials, which tremendously broadens the range of valley materials.

\begin{figure}[htbp]
\centering
\flushleft
\includegraphics[width=0.9\linewidth,clip]{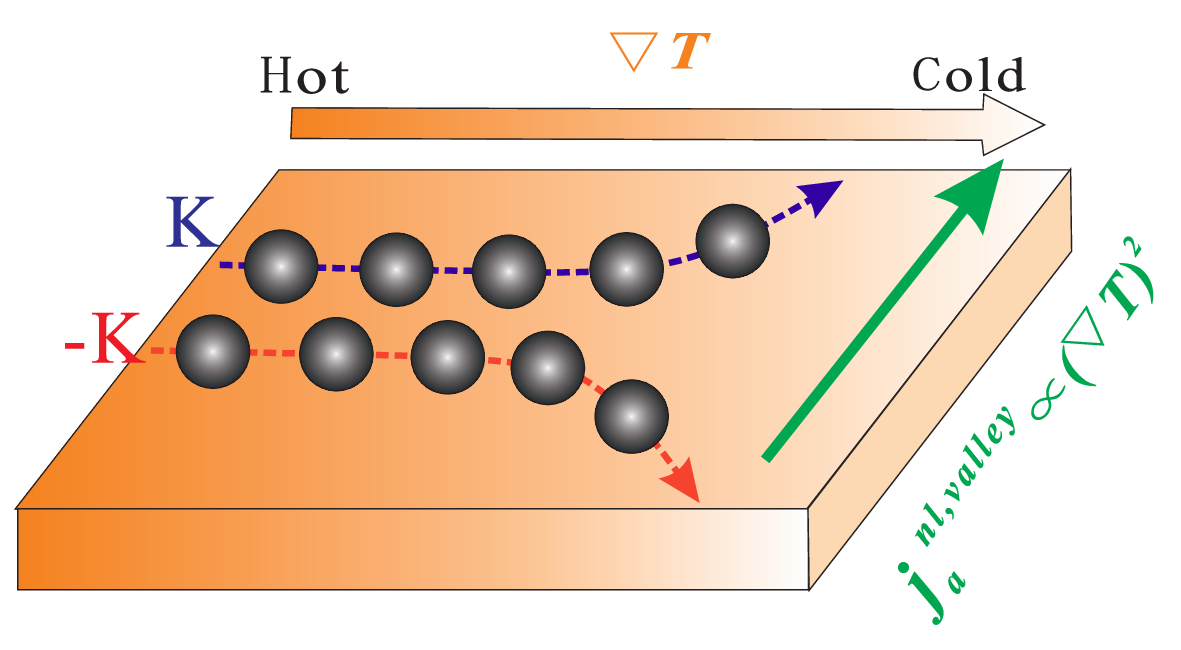}
\caption{Schematic of the nonlinear valley Nernst effect. A transverse nonlinear valley Nernst current $\propto (\nabla T)^{2}$ is generated as a second-order response to the longitudinal temperature gradient in $\mathcal{P}-$ and $\mathcal{T}-$ symmetric materials, where the linear valley current and transverse linear (nonlinear) charge Nernst current disappear.}
\label{figure1}
\end{figure}

In this paper, we theoretically explore the thermally driven nonlinear valley Nernst effect (NVNE) [Fig.~\ref{figure1}] as a second-order response to the temperature gradient through the semiclassical theory in the nonmagnetic centrosymmetric materials, in  which both $\mathcal{P}$ and $\mathcal{T}$ symmetries are present. Additionally, we also systematically analyze both global and local symmetry constraints on NVNE. Our study shows that the intrinsic NVNE (independent of $\tau$) originating from the quantum metric can exist in materials with both global $\mathcal{T}$ and $\mathcal{P}$ symmetries, and the largest local symmetry near the valleys in 2D systems for a nonvanishing intrinsic NVNE is a single mirror symmetry. 
The formulas of the nonlinear thermoelectric coefficients (NTCs) originating from the intrinsic, BC, and Drude contributions are determined, and the symmetry constraints on the thermally driven linear and nonlinear valley (charge) currents are analyzed and discussed in Sec.~\ref{NVNE}. A correction stemming from the temperature gradient to the orbital magnetic moment (OMM) is introduced and discussed in Sec.~\ref{TGIOM}. The effective Hamiltonian of the strained gapless trigonal-warping bilayer graphene is given, and its symmetry is analyzed in Sec.~\ref{EHSA}. The behaviours of the intrinsic NVNE for the strained bilayer graphene are discussed in Sec.\ref{ReDi}. Finally, a conclusion is given in Sec.~\ref{conclusion}.

\section{THEORETICAL DERIVATION and analysis}
\label{NVNE}
It has been established that the linear anomalous valley Nernst current $j_{a}^{\mathrm{valley}}\equiv\alpha_{ab}^{\mathrm{valley}}\partial_{b}T$, where $j_{a}^{\mathrm{valley}}=j_{a}^{\mathrm{K}}-j_{a}^{\mathrm{K}^{\prime}}$ and  $\alpha_{ab}^{\mathrm{valley}}=\alpha_{ab}^{\mathrm{K}}-\alpha_{ab}^{\mathrm{K}^{\prime}}$ with the distinct valleys specified by $\mathrm{K}$ and $\mathrm{K}^{\prime}$, can't exist in  $\mathcal{P}$-symmetric systems with $\mathcal{T}$ symmetry. That's because the linear anomalous valley Nernst coefficient $\alpha_{ab}^{\mathrm{valley}}$ is restricted to be even under $\mathcal{T}$ symmetry but odd under $\mathcal{P}$ symmetry since the valley current $j_{a}^{\mathrm{valley}}$ keeps unchanged under $\mathcal{T}$ or $\mathcal{P}$ operation ($j_{a}^{\mathrm{K}}\xrightarrow{\mathcal{T}\,\text{or}\,\mathcal{P}} -j_{a}^{\mathrm{K}^{\prime}}$, $j_{a}^{\mathrm{K}^{\prime}}\xrightarrow{\mathcal{T}\,\text{or}\,\mathcal{P}} -j_{a}^{\mathrm{K}}$) while the temperature gradient $\partial_{b}T$ remains unchanged under $\mathcal{T}$ but changes sign under $\mathcal{P}$. As a result, $\alpha_{ab}^{\mathrm{valley}}$ is forced to be zero in presence of both $\mathcal{P}$ and $\mathcal{T}$ symmetries, meaning  $j_{a}^{\mathrm{valley}}=0$. 
In this background, our sight turns into the nonlinear response and investigate the nonlinear valley Nernst current $j_{a}^{\mathrm{nl,valley}}$ (where the superscript ``{nl}" represents nonlinear) whose concept roots in the nonlinear charge thermoelectric response specified by $j_{a}^\text{nl}=\alpha_{abc}\partial_{b}T\partial_{c}T$. $j_{a}^{\mathrm{nl,valley}}$ is defined as $j_{a}^{\mathrm{nl,valley}}=\alpha_{abc}^{\mathrm{valley}}\partial_{b}T\partial_{c}T$ with
\begin{equation}
\alpha_{abc}^{\mathrm{valley}}=\alpha_{abc}^{\mathrm{K}}-\alpha_{abc}^{\mathrm{K}^{\prime}}.
\end{equation}
Based on the dependence on the scale of the relaxation time $\tau$, we decompose the nonlinear thermoelectric currents ${j}^\text{nl}_{a}$ in $a$ direction into three components as
 ${j}^{\mathrm{nl}}_{a} ={j}^{\mathrm{nl}}_{1,a}(\tau^{0})+{j}^{\mathrm{nl}}_{2,a}(\tau^{1})+
{j}^{\mathrm{nl}}_{3,a}(\tau^{2})$ with ${j}^{\mathrm{nl}}_{i(=1,2,3),a}=\alpha_{abc,i}\partial_{b}T\partial_{c}T$.
The NTCs  $\alpha_{abc,1}\propto \tau^{0}$, $\alpha_{abc,2} \propto \tau$ and $\alpha_{abc,3} \propto \tau^{2}$ are disclosed to originate from the quantum metric, BC, and Drude contribution, respectively (see the details in Appendix \ref{Ap-A}). Therefore, we rewritten the coefficients [$\alpha_{abc,1}$, $\alpha_{abc,2}$, $\alpha_{abc,3}$] as [$\alpha^{\text{in}}_{abc}$, $\alpha^{\text{B}}_{abc}$, $\alpha^{\text{D}}_{abc}$], where superscript ``{in}",  ``{B}", and  ``{D}"  represents intrinsic, Berry, and Drude, respectively. 
After a tedious derivation in Appendix \ref{Ap-A},
the NTCs [$\alpha^{\text{in}}_{abc}$, $\alpha^{\text{B}}_{abc}$, $\alpha^{\text{D}}_{abc}$] are determined as, respectively,
\begin{eqnarray}
\alpha_{abc}^{\mathrm{in}}&=&\frac{e}{\hbar}\sum_{n}\int[d\boldsymbol{k}]
\frac{(\varepsilon _{\boldsymbol{k}}^{n}-E_{f})^{2}}{T^{2}}\frac{\partial f_{0}}{\partial \varepsilon _{\boldsymbol{k}}^{n}}\chi_{abc}^{n}(\boldsymbol{k}),\label{tau-0}\\
\alpha_{abc}^{\mathrm{B}}&=&-\frac{e\tau}{\hbar} \epsilon_{abd} \sum_{n} \int[d\boldsymbol{k}]
 \frac{(\varepsilon ^{n}_{\boldsymbol{k}}-E_{f})^{2}}{T^{2}}
v_{c}^{n}\frac{\partial f_{0}}{\partial \varepsilon ^{n}_{\boldsymbol{k}}}\Omega^{n}_{d}(\boldsymbol{k}),\label{tau-1}\\
\alpha_{abc}^{\mathrm{D}}&=&\frac{e\tau^{2}}{T^{2}}\sum_{n}\int[d\boldsymbol{k}]\frac{(\varepsilon ^{n}_
{\boldsymbol{k}}-E_{f})^{2}}{m_{ab}({\boldsymbol{k}})}v_{c}^{n}\frac{\partial f_{0}}{\partial \varepsilon ^{n}_{\boldsymbol{k}}},\label{tau-2}
\end{eqnarray}
where $\int[d\boldsymbol{k}]$ is shorthand for $\int d\boldsymbol{k}/(2\pi)^{d}$ with $d$ denoting the dimension of the system,
$f_{0}$ is the equilibrium Fermi-Dirac distribution function,
 $\epsilon_{abd}$ presents the Levi-Civita symbol, $\boldsymbol{v}^{n}(\boldsymbol{k})=\langle u_{n}(\boldsymbol{k})|\hat{\boldsymbol{v}}|u_{n}(\boldsymbol{k})\rangle$ indicates the intraband velocity matrix element for $n$th band determining by the periodic part of the Bloch wave-function  $|u_{n}(\boldsymbol{k})\rangle$, $\varepsilon _{\boldsymbol{k}}^{n}$ denotes  the $n$th-energy band for the unperturbed Hamiltonian, $\boldsymbol{\mathcal{A}}^{nm}(\boldsymbol{k})=\langle u_{n}(\boldsymbol{k})|i\nabla_{\boldsymbol{k}}|u_{m}(\boldsymbol{k})\rangle$ denotes the interband Berry connection,  $\Omega^{n}_{a}(\boldsymbol{k})=\sum_{m\neq n}\epsilon_{abc}\Omega_{bc}^{nm}/2$ gives the  BC for the $n$-th band with $\Omega_{bc}^{nm}(\boldsymbol{k})=i(\mathcal{A}_{b}^{nm}\mathcal{A}_{c}^{mn}-\mathcal{A}_{c}^{nm}\mathcal{A}_{b}^{mn})$ representing the band-resolved BC \cite{H. Watanabe},
$m_{ab}^{-1}(\boldsymbol{k})=(1/\hbar)(\partial{v}_{a}^{n}/\partial{k_{b}})$ shows the inverse effective mass tensor, and the coefficient $\chi_{abc}^{n}(\boldsymbol{k})$ is defined as  
\begin{equation}
\begin{aligned}
\chi_{abc}^{n}(\boldsymbol{k})=\sum_{m\neq n}\hbar\left[\frac{2v_{a}^{n}\mathcal{G}_{bc}^{mn}-v_{b}^{n}\mathcal{G}_{ac}^{mn}-v_{c}^{n}
\mathcal{G}_{ab}^{mn}}{(\varepsilon _{\boldsymbol{k}}^{n}-\varepsilon _{\boldsymbol{k}}^{m})}\right],
\end{aligned}
\label{CHI-COE}
\end{equation}
with $\mathcal{G}_{ab}^{mn}(\boldsymbol{k})=\mathrm{Re}[\mathcal{A}_{a}^{nm}\mathcal{A}_{b}^{mn}]$ indicting the band-resolved quantum metric\cite{Wang1,Zhuang,N.-Z. Wang,A. Gao}.
The presence of the factor $\partial f_{0}/\partial \varepsilon _{\boldsymbol{k}}^{n}$ in Eqs.~\eqref{tau-0}-\eqref{tau-2} hints that the thermal-driven nonlinear current is a Fermi surface feature.

\begin{table*}[tbph]
\centering
\caption{Constraints on the in-plane tensor elements of $\alpha^{\mathrm{in,valley}}_{abc}$ from point group symmetries. $'\checkmark'$ ($'\times'$) means the element is symmetry allowed (forbidden).}
\begin{centering}
\begin{tabular*}{18 cm}{@{\extracolsep{\fill}}ccccc ccccc ccccc}
\hline \hline
   \multicolumn{1}{c} { } & \multicolumn{2}{c} {global }     &\multicolumn{11}{c} {local  }\\
   \cline{2-3}\cline{4-15}
                          &  $\mathcal{P}$ & $\mathcal{T}$      &  $\mathcal{P}$
                           & $C_{n(n>2)}^{z}$  &  $C_{n(n>2)}^{x}$    &  $C_{n(n>2)}^{y}$  &  $M_{z}$       &    $M_{x}$     &    $M_{y}$
                          &  $S_{4,6}^{z} $        &  $S_{4}^{x}$       &  $S_{6}^{x}$     &  $S_{4}^{y}$       &  $S_{6}^{y}$ \\
   \cline{1-1} \cline{2-3}\cline{4-15}
$\alpha^{\mathrm{in,valley}}_{yxx}$ & $ \checkmark$  & $ \checkmark$  &  $ \times  $
                         &$ \times  $          &  $ \times  $     & $\checkmark$       &   $ \checkmark$     &   $ \checkmark$        &    $\times  $
                         &    $\times$            &      $\times$        &   $ \times  $    &    $ \checkmark$     &   $ \times  $\\
$\alpha^{\mathrm{in,valley}}_{xyy}$ & $ \checkmark$  &  $ \checkmark$   &   $ \times  $
                          & $ \times  $          &   $ \checkmark$  & $ \times$      & $ \checkmark$       &     $\times  $        &   $ \checkmark$
                         &     $\times$           &    $ \checkmark$     &   $ \times  $   &      $\times$        &   $ \times  $ \\
 \hline \hline
\end{tabular*}
\par\end{centering}
\label{table1}
\end{table*}

Under $\mathcal{T}$ symmetry, the current $j^\text{nl}_{i,a}$ and $\tau$ change sign but $\partial_{b}T\partial_{c}T$ keeps unchanged, restricting  the $\tau$-independent part of $\alpha_{abc}^{\mathrm{B}}$ $\mathcal{T}$-even  ($\alpha_{abc}^{\mathrm{B}}/\tau\xrightarrow{\mathcal{T}}\alpha_{abc}^{\mathrm{B}}/\tau$) while the $\tau$-independent part of $\alpha_{abc}^{\mathrm{in}}$ and $\alpha_{abc}^{\mathrm{D}}$ $\mathcal{T}$-odd ($\alpha_{abc}^{\mathrm{in}}\xrightarrow{\mathcal{T}}-\alpha_{abc}^{\mathrm{in}}$ and $\alpha_{abc}^{\mathrm{D}}/\tau^{2}\xrightarrow{\mathcal{T}}-\alpha_{abc}^{\mathrm{D}}/\tau^{2}$).
These $\mathcal{T}$-odd parities force $\alpha_{abc}^{\mathrm{in}}$ and $\alpha_{abc}^{\mathrm{D}}$ to vanish in the $\mathcal{T}$-symmetric systems.
 In fact, the disappearance of coefficients [$\alpha_{abc}^{\mathrm{in}}$ and $\alpha_{abc}^{\mathrm{D}}$ ] in $\mathcal{T}$-symmetric systems can also be confirmed through exploiting the constraints on the quantities [$\varepsilon ^{n}(\boldsymbol{k})$, $\boldsymbol{v}^{n}\left(\boldsymbol{k}\right)$, $\Omega^{nm}_{ab}(\boldsymbol{k})$, $\mathcal{G}_{ab}^{mn}(\boldsymbol{k})$, $m^{-1}_{ab}(\boldsymbol{k})$] under $\mathcal{T}$ symmetry given in Table $\ref{App-B-table}$. According to those constraints, one can easily identify the integrands in Eqs. $\eqref{tau-0}$ and $\eqref{tau-2}$ are odd functions of $\boldsymbol{k}$ when system is $\mathcal{T}$ symmetric, resulting in $\alpha_{abc}^{\mathrm{in}}=0$ and $\alpha_{abc}^{\mathrm{D}}=0$. Besides, when in presence of both the $\mathcal{T}$ and $\mathcal{P}$ symmetries, the coefficient $\alpha_{abc}^{\mathrm{B}}$ also becomes zero. That's because the BC satisfies $\Omega_{ab}^{nm}(\boldsymbol{k})= -\Omega_{ab}^{nm}(-\boldsymbol{k})$ under $\mathcal{T}$ symmetry and $\Omega_{ab}^{nm}(\boldsymbol{k})= \Omega_{ab}^{nm}(-\boldsymbol{k})$ under $\mathcal{P}$ symmetry. Subsequently, when both $\mathcal{T}$ and $\mathcal{P}$ symmetries are present, $\Omega_{ab}^{nm}(\boldsymbol{k})=0$ at each point in the momentum space, which leads to $\alpha^{\mathrm{B},\mathrm{K}}_{abc}=0$ and $\alpha^{\mathrm{B},\mathrm{K}^{\prime}}_{abc}=0$, hinting both thermal-driven nonlinear valley and charge currents stemming from the BC near the Fermi surface disappear. Therefore, the nonlinear thermal-driven current disappears in the $\mathcal{P}$-symmetric systems with $\mathcal{T}$ symmetry.

Be different to the BC, the quantum metric $\mathcal{G}_{ab}^{mn}(\boldsymbol{k})$ can exist in presence of both $\mathcal{P}$ and $\mathcal{T}$ symmetries since the sign of $\mathcal{G}_{ab}^{mn}(\boldsymbol{k})$ keeps unchange whether under $\mathcal{T}$ or $\mathcal{P}$ symmetry, namely $\mathcal{G}_{ab}^{mn}(\boldsymbol{k})\xrightarrow{\mathcal{T}/\mathcal{P}}\mathcal{G}_{ab}^{mn}(-\boldsymbol{k})$.
Hence, both the quantum metric and Drude contribution might contribute to the thermally driven nonlinear valley current in the systems with both $\mathcal{T}$ and $\mathcal{P}$ symmetries.
In the following, we focus on the NVNE arising from the quantum metric and exclude the Drude contribution. Experimentally, the signal from the Drude contribution can be separated based on the scaling relations involving the linear conductivity $\sigma\propto\tau$ : Drude contribution (quadratic to $\sigma$) and quantum metric (independent of $\sigma^{2}$).



In addition to the constraints from the $\mathcal{T}$ and $\mathcal{P}$ symmetries, the other local symmetries near the individual valleys will also impose constraints on the nonlinear coefficients $\alpha_{a^{\prime}b^{\prime}c^{\prime}}$ for $\mathrm{K}$ ($\mathrm{K}^{\prime}$) valleys as:
\begin{equation}
\alpha^{\mathrm{K}/\mathrm{K}^{\prime}}_{a^{\prime}b^{\prime}c^{\prime}}
=\mathcal{R}_{a^{\prime}a}\mathcal{R}_{b^{\prime}b}\mathcal{R}_{c^{\prime}c}
\alpha^{\mathrm{K}/\mathrm{K}^{\prime}}_{abc},
\end{equation}
where $\mathcal{R}$ is a point group operation. 
Here, it should be point out that $\mathcal{R}$ indicates symmetries of the low-energy $k\cdot p$ model defined for the valleys instead of the global crystalline symmetry since the local symmetries near the valleys determine the valley response instead of global crystalline symmetry\cite{Das Kamal}. The obtained constraints on coefficients [$\alpha^{\mathrm{in,valley}}_{xyy}$, $\alpha^{\mathrm{in,valley}}_{yxx}$] from the intrinsic contribution
are summarized in Table \ref{table1}. According to Table \ref{table1}, one can easily notice that the largest local symmetry near the valleys in 2D cases  for a nonvanishing $\alpha^{\mathrm{in,valley}}_{aa_{\perp}a_{\perp}}$ is a single mirror symmetry $M_{a_{\perp}}$, where $a_{\perp}$ indicates the coordinate axis orthogonal to axis $a$ in the 2D plane.

\section{Temperature-gradient-induced orbital magnetic moment}\label{TGIOM}
In linear VNE, the different transverse edges will hold the oppositely OMM polarized carriers in the real space which is guaranteed by the valley contrasting OMM, i.e. $\boldsymbol{m}^{\mathrm{K}}(\boldsymbol{k})=-\boldsymbol{m}^{\mathrm{K}^{\prime}}(\boldsymbol{k})$. One might ask what physical quantity distinguishes the carriers separated by the intrinsic NVNE in real space since the OMM vanishes ($\boldsymbol{m}(\boldsymbol{k})=0$) in the whole Brillouin zone when both $\mathcal{P}$ and $\mathcal{T}$ symmetries are present.

We notice that an correction to the OMM will be induced by the temperature gradient and can be  nonzero when in presence of both $\mathcal{P}$ and $\mathcal{T}$ symmetries. The $a$-th component of this temperature-gradient induced OMM $m_{n,a}^{\nabla T}(\boldsymbol{k})$  for $n$ band is \cite{C. Xiao1,C. Xiao2}
\begin{equation}
\begin{aligned}
m_{n,a}^{\nabla T}(\boldsymbol{k})=\sum_{  m\neq n     }
\left[2\mathrm{Re}\frac{\mathcal{M}_{a}^{ nm  }\mathcal{A}_{d}^{ mn}}{\varepsilon ^{n}_{\boldsymbol{k}}-\varepsilon ^{m}_{\boldsymbol{k}}}
      +\frac{e}{2\hbar}\epsilon_{abc}(\partial_{b}\mathcal{G}_{cd}^{mn }) \right]\partial_{d} T,
\end{aligned}
\label{OMM}
\end{equation}
with   $\boldsymbol{\mathcal{M}}^{nm}=\frac{e}{2}\sum_{l\neq m}(\boldsymbol{v}^{nl}+\boldsymbol{v}^{m}\delta_{nl})\times\boldsymbol{\mathcal{A}}^{lm}$ representing the interband OMM,
where $\boldsymbol{v}^{nl}(\boldsymbol{k})=\langle u_{n}(\boldsymbol{k})|\hat{\boldsymbol{v}}|u_{l}(\boldsymbol{k})\rangle$ denotes the interband velocity matrix element.

Through exploiting the constraints on the quantities [$\partial_{d}T$, $\varepsilon ^{n}(\boldsymbol{k})$, $\boldsymbol{\mathcal{A}}^{nm}(\boldsymbol{k})$, $\boldsymbol{v}^{n}\left(\boldsymbol{k}\right)$, $\mathcal{G}_{ab}^{mn}(\boldsymbol{k})$] under $\mathcal{P}/\mathcal{T}$ symmetry given in Table $\ref{App-B-table}$, one can easily verify that the temperature-gradient induced OMM $\boldsymbol{m}_{n}^{\nabla T}(\boldsymbol{k})$ changes signs under both $\mathcal{P}$ and $\mathcal{T}$ symmetries, namely $\boldsymbol{m}_{n}^{\nabla T}(\boldsymbol{k})\xrightarrow{\mathcal{T}/\mathcal{P}}-\boldsymbol{m}_{n}^{\nabla T}(-\boldsymbol{k})$, which hints 1) $\boldsymbol{m}_{n}^{\nabla T}(\boldsymbol{k})$ can exist when both $\mathcal{P}$ and $\mathcal{T}$ symmetries are present; 2) the nonlinear OMM satisfies
$\boldsymbol{m}_{n}^{\nabla T,\mathrm{K}}(\boldsymbol{k})=-\boldsymbol{m}_{n}^{\nabla T,\mathrm{K}^{\prime}}(-\boldsymbol{k})$. 
 \begin{figure*}[htbp]
\centering
\flushleft
\includegraphics[width=1.0\linewidth,clip]{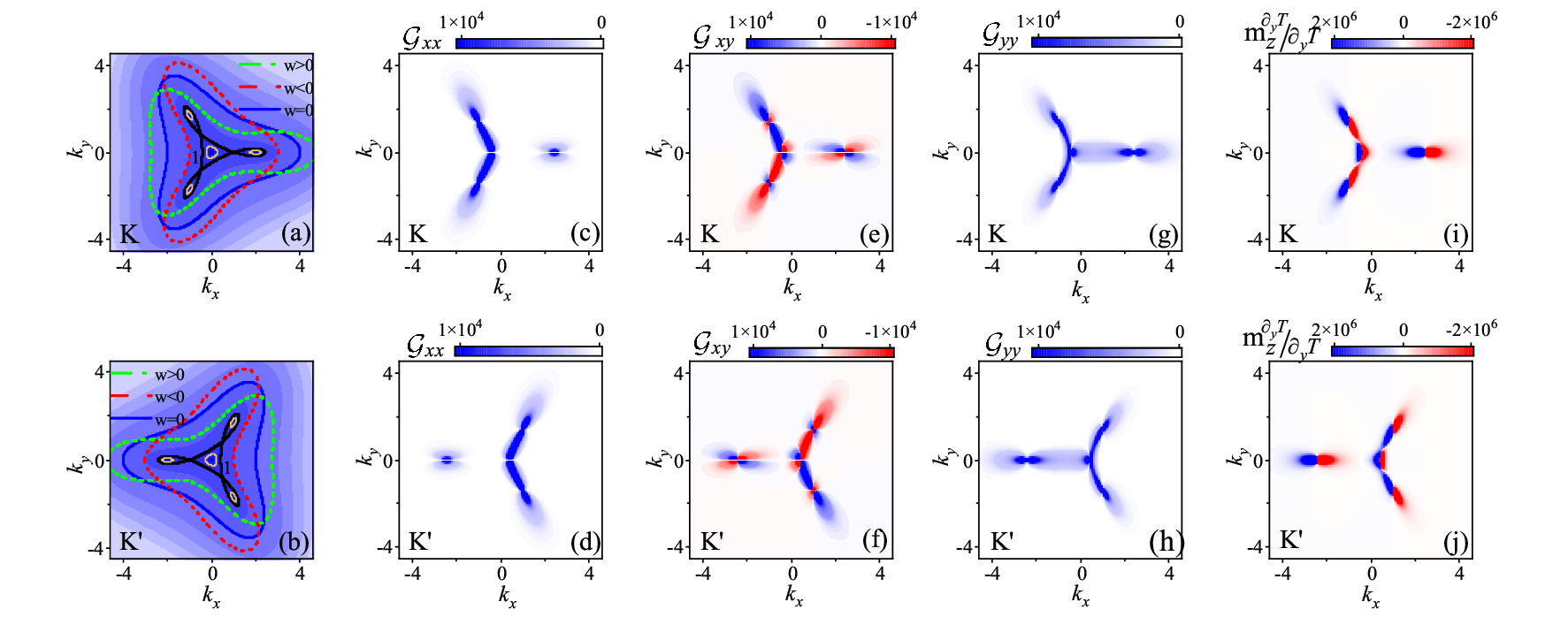}
\caption{  Schematic of the energy contour [(a),(b)], the quantum metric elements $\mathcal{G}_{xx}$ [(c),(d)], $\mathcal{G}_{xy}$ [(e),(f)], $\mathcal{G}_{yy}$[(g),(h)]
          and the temperature-gradient-induced orbital magnetic moment $m_{z}^{\partial_{y} T}$ [(i), (j)] of the conduction band for different valleys of trigonal warping bilayer graphene. The up (bottom) ones are for $\mathrm{K}$ ($\mathrm{K}^{\prime}$) valleys, respectively. The green dash dot line (red dash line) and the solid lines in [(a),(b)] indicate energy band with the uniaxial tensile (compressive) strain and without strain, respectively. The color background in (a) and (b) represent the energy contour without strain ($w=0$). The strength of strain $w=1\varepsilon_{L}$ is fixed in (c)-(j). Momenta (energies) are measured in units of $k_{L}=m^{*}v_{3}/\hbar$  ($\varepsilon_{L}=m^{*}v_{3}^{2}/2$). The $\mathcal{G}_{ab}$ and  $m_{z}^{\partial_{y} T}/\partial_{y} T$ are measured in units of $k_{L}^{-2}$ and $e/(\hbar k_{L}^{3})$, respectively.
            }
\label{fig2}
\end{figure*}
Additionally, the entropy density $s^{\tau_{v}}_{n0}(\boldsymbol{k})=f_{0}(\varepsilon^{n}_{\boldsymbol{k}}-E_{f})/T+k_{B}T\ln[1+
\exp(-\frac{\varepsilon^{n}_{\boldsymbol{k}}-E_{f}}{k_{B}T})]$ for opposite valleys meets $s_{n0}^{\mathrm{K}}(\boldsymbol{k})=s_{n0}^{\mathrm{K}^{\prime}}(-\boldsymbol{k})$ in the $\mathcal{T}-$ and $\mathcal{P}$-symmetric systems. As a result, the orbital magnetization  $\boldsymbol{M}_{n}^{\nabla T,\tau_{v}}=\int_{\boldsymbol{K}_{\tau_{v}}}[d\boldsymbol{k}]s^{\tau_{v}}_{n0}(\boldsymbol{k})
\boldsymbol{m}_{n}^{\nabla T,\tau_{v}}(\boldsymbol{k})$ for the $\mathrm{K}$ valley, which stems from the temperature-gradient-induced OMM, has same magnitude but opposite signs to that for the $\mathrm{K}^{\prime}$ valley (i.e. $\boldsymbol{M}_{n}^{\nabla T,\mathrm{K}}=-\boldsymbol{M}_{n}^{\nabla T,\mathrm{K}^{\prime}}$), meaning that the temperature-gradient-induced orbital magnetization is  valley-contrasting.
 Here $\int_{\boldsymbol{K}_{\tau_{v}}}$ represents the integration near $\tau_{v}$ valley.
Hence, when applying a temperature gradient to the materials exhibiting the two fundamental symmetries ($\mathcal{T}$ and $\mathcal{P}$ symmetries), the carriers in the inequivalent valleys hold opposite polarizations of orbital magnetization and flow towards opposite transverse edges in real space driven  by the NVNE, hinting the intrinsic valley-contrasting orbital magnetization (VCOM) accompanies the intrinsic NVNE and distinguishes the carriers separated by the intrinsic NVNE in real space. Owing to the accompaniment of VCOM, one would detect the NVNE through the magneto-optical Kerr method\cite{J. Lee2016,J. Son2019} (see the details in Appendix \ref{App-C}).

\section{The effective Hamiltonian and symmetries analysis for the strained bilayer graphene}\label{EHSA}
One of candidate materials to observe the intrinsic NVNE is the strained  trigonal-warping bilayer graphene in the $AB$ Bernal-stacked structure. 
Both global $\mathcal{P}$ and $\mathcal{T}$ symmetries are protected in the $AB$ Bernal-stacked bilayer graphene since its crystal point symmetry belongs to $D_{3d}$\cite{J. L. Maes,Sylvain Latil,Mikito Koshino}. Additionally, the local $\mathcal{P}$ symmetry is lacking in the $AB$ Bernal-stacked bilayer graphene near two valleys [Figs.~\ref{fig2} (a) and (b)] since its local symmetry belongs to $C_{3v}$ (comprised of a threefold rotation $C_{3}$ and a mirror symmetry) instead of $C_{6v}$ owing to the interlayer skew hopping. One might notice that the local $C_{3}$ symmetry near the valleys would force the intrinsic NVNE vanishing in $AB$ Bernal-stacked bilayer graphene (Table~\ref{table1}). However, the local $C_{3}$ symmetry can be broken through the application of a uniaxial strain, resulting in nonzero intrinsic NVNE.
This uniaxial strain would be experimentally induced in bilayer graphene through mechanical bending, typically achieved  by placing it on flexible substrates such as polymethyl methacrylate (PMMA)\cite{J. Son2019}, polydimethylsiloxane (PDMS)\cite{M. Huang2009}, or polyethylene terephthalate (PET)\cite{T. Yu2008} film.

In the presence of uniaxial strain along $x$ direction, the effective low-energy Hamiltonian for $\tau_{v}$ ($=\pm1$, $+1$ for $\mathrm{K}$ and $-1$ for $\mathrm{K}^{\prime}$) valley of the trigonal-warping bilayer graphene\cite{E. McCann1,M. Mucha-Kruczynski} can be expressed as
\begin{equation}
\begin{aligned}
\hat{H}_{\tau_{v}}& =\left(-\frac{\hbar^{2}(k_{x}^{2}-k_{y}^{2})}{2m^{*}}+w \right)\sigma_{x}  -\tau_{v}\frac{\hbar^{2}k_{x}k_{y}}{m^{*}} \sigma_{y}\\
&+\tau_{v}v_{3}\hbar k_{x}\sigma_{x}-v_{3}\hbar k_{y}\sigma_{y},
\end{aligned}
\label{Ham}
\end{equation}
where $m^{*}$ is the effective mass directly dependent on the interlayer coupling, $w$ is attributed to the strain effect and depends on the strain tensors $u_{xx}$ and $u_{yy}$ as $w=\zeta(u_{xx}-u_{yy})$, with $\zeta=-0.443$ for the bilayer graphene \cite{M. Mucha-Kruczynski}, the terms in the second line originate from the interlayer skew hopping and induce a trigonal warping deformation of the Fermi circle near the Dirac cone [Figs.~\ref{fig2} (a) and (b)], $v_{3}$ indicates the Fermi velocity related to the interlayer skew hopping, and $\sigma_{i=x,y,z}$ represents the pauli matrix for sublattices. For simplicity, we focus on the conduction band, and the corresponding energy dispersion $\varepsilon _{\tau_{v}}(\boldsymbol{k})$, the intraband velocity matrix element $v_{\tau_{v},a}(\boldsymbol{k})$, and the quantum metric $\mathcal{G}_{\tau_{v},ab}(\boldsymbol{k})$ of the conduction band are found to be, respectively
\begin{equation}
\begin{aligned}
&\varepsilon _{\tau_{v}}(\boldsymbol{k})=\sqrt{N_{1}^{2}+N_{2}^{2}}, \,\,\,v_{\tau_{v},a}(\boldsymbol{k})=\frac{N_{1}N_{1a}+N_{2}N_{2a}}{\sqrt{N_{1}^{2}+N_{2}^{2}}},\\
&\mathcal{G}_{\tau_{v},ab}(\boldsymbol{k})=  \\
              &   \frac{ N_{2}^{2}N_{1a}N_{1b}+N_{1}^{2}N_{2a}N_{2b}-N_{1}N_{2}(N_{1a}N_{2b}+N_{1b}N_{2a}) }{4(N_{1}^{2}+N_{2}^{2})^{2}} ,\\
\end{aligned}
\end{equation}
where  $N_{1}(\boldsymbol{k})=-\hbar^{2}(k_{x}^{2}-k_{y}^{2})/2m^{*}+\tau_{v}v_{3}\hbar k_{x}+w$, $N_{2}(\boldsymbol{k})=-\tau_{v}\hbar^{2}k_{x}k_{y}/m^{*}-v_{3}\hbar k_{y}$, the variable $\boldsymbol{k}$ in bracket of $N_{i(=1,2)}(\boldsymbol{k})$ and $N_{ij}(\boldsymbol{k})=\partial[N_{i}(\boldsymbol{k})]/\partial k_{j=a,b}$ has been neglected for simplification, and the energy-band label ``{c}" representing conduction band in the superscripts of physical quantities [$\varepsilon $, $v$ and $\mathcal{G}$] have also been neglected for simplicity. As expected, the quantum metric elements [$\mathcal{G}_{xx}$, $\mathcal{G}_{xy}$, and $\mathcal{G}_{yy}$] are nonzero and satisfy $\mathcal{G}^{\mathrm{K}}_{ab}(\boldsymbol{k})=\mathcal{G}^{\mathrm{K}^{\prime}}_{ab}(-\boldsymbol{k})$ [Figs.~\ref{fig2}(c)-(h)] in presence of $\mathcal{P}$ and $\mathcal{T}$ symmetries.


It should be pointed out that the effective Hamiltonian $\hat{H}_{\tau_{v}=+1}$  and $\hat{H}_{\tau_{v}=-1}$ in Eq.~\eqref{Ham} act in the same space of two-component wave functions $\Phi=(\phi_{A},\phi_{\tilde{B}})$, where $\phi_{\alpha (=A,\tilde{B})}$ is the electron amplitude on the sublattice $\alpha$. This framework differs from the previous works\cite{E. McCann1,M. Mucha-Kruczynski}, where the component in $\Phi_{\tau_{v}=-1}$ is reversed for valley $\mathrm{K}^{\prime}$, namely $\Phi_{\tau_{v}=-1}=(\phi_{\tilde{B}},\phi_{A})$. In this space  $\Phi=(\phi_{A},\phi_{\tilde{B}})$, we have the time-reversal operator $\hat{\mathcal{T}}=\sigma_{0}\mathcal{K}$ with $\mathcal{K}$ representing complex conjugation operator, inversion operator $\hat{\mathcal{P}}=\sigma_{x}$, the mirror symmetries $M_{x}=\sigma_{0}$ and $M_{y}=\sigma_{x}$. 
Consequently, one can easily confirm that both global $\mathcal{P}$ and  $\mathcal{T}$ symmetries are present in the strained trigonal-warping bilayer graphene since $\hat{\mathcal{T}} \hat{H}_{\mathrm{K}}(\boldsymbol{k})\hat{\mathcal{T}}^{-1}=\hat{H}_{\mathrm{K}^{\prime}}
(-\boldsymbol{k})$ and $\hat{\mathcal{P}}\hat{H}_{\mathrm{K}}(\boldsymbol{k})\hat{\mathcal{P}}^{-1}=
\hat{H}_{\mathrm{K}^{\prime}}(-\boldsymbol{k})$. On the contrary, the local $\mathcal{P}$, $\mathcal{T}$, and $M_{x}$ symmetries are broken, and only the local $M_{y}$ symmetry is survived owing to $\hat{\mathcal{T}} \hat{H}_{\tau_{v}}(\boldsymbol{k})\hat{\mathcal{T}}^{-1}\neq \hat{H}_{\tau_{v}}(-\boldsymbol{k})$, $\hat{\mathcal{P}}\hat{H}_{\tau_{v}}(\boldsymbol{k})\hat{\mathcal{P}}^{-1}\neq
 \hat{H}_{\tau_{v}}(-\boldsymbol{k})$,  $\hat{M}_{x}\hat{H}_{\tau_{v}}(k_{x},k_{y})\hat{M}_{x}^{-1}\neq
 \hat{H}_{\tau_{v}}(-k_{x},k_{y})$ and $\hat{M}_{y}\hat{H}_{\tau_{v}}(k_{x},k_{y})\hat{M}_{y}^{-1}=
 \hat{H}_{\tau_{v}}(k_{x},-k_{y})$.
Additionally, owing to the presence of the uniaxial strain ($w\neq0$), the local threefold rotation $C_3$ is also broken since $\hat{C}_{3}\hat{H}_{\tau_{v}}(\hat{R}_{2\pi/3}\boldsymbol{k})\hat{C}_{3}^{-1}\neq\hat{H}_{\tau_{v}}
(\boldsymbol{k})$ and $\hat{C}^{2}_{3}\hat{H}_{\tau_{v}}(\hat{R}_{-2\pi/3}\boldsymbol{k})(\hat{C}_{3}^{2})^{-1}
\neq\hat{H}_{\tau_{v}}(\boldsymbol{k})$, where the threefold rotation operator for the $\tau_{v}$ valley is expressed as $C_{3}=e^{\frac{i2\pi}{3}\tau_{v}\sigma_{z}}$ and $C^{2}_{3}=e^{-\frac{i2\pi}{3}\tau_{v}\sigma_{z}}$, and $\hat{R}_{\theta}$ represents the rotation operator around $z$-axis by $\theta(=\pm2\pi/3)$. As a result, only the local mirror symmetry $M_{y}$ is survived in the trigonal-warping bilayer graphene in presence of the uniaxial strain along $x$-direction.

 \begin{figure}[htbp]
\centering
\flushleft
\includegraphics[width=1.0\linewidth,clip]{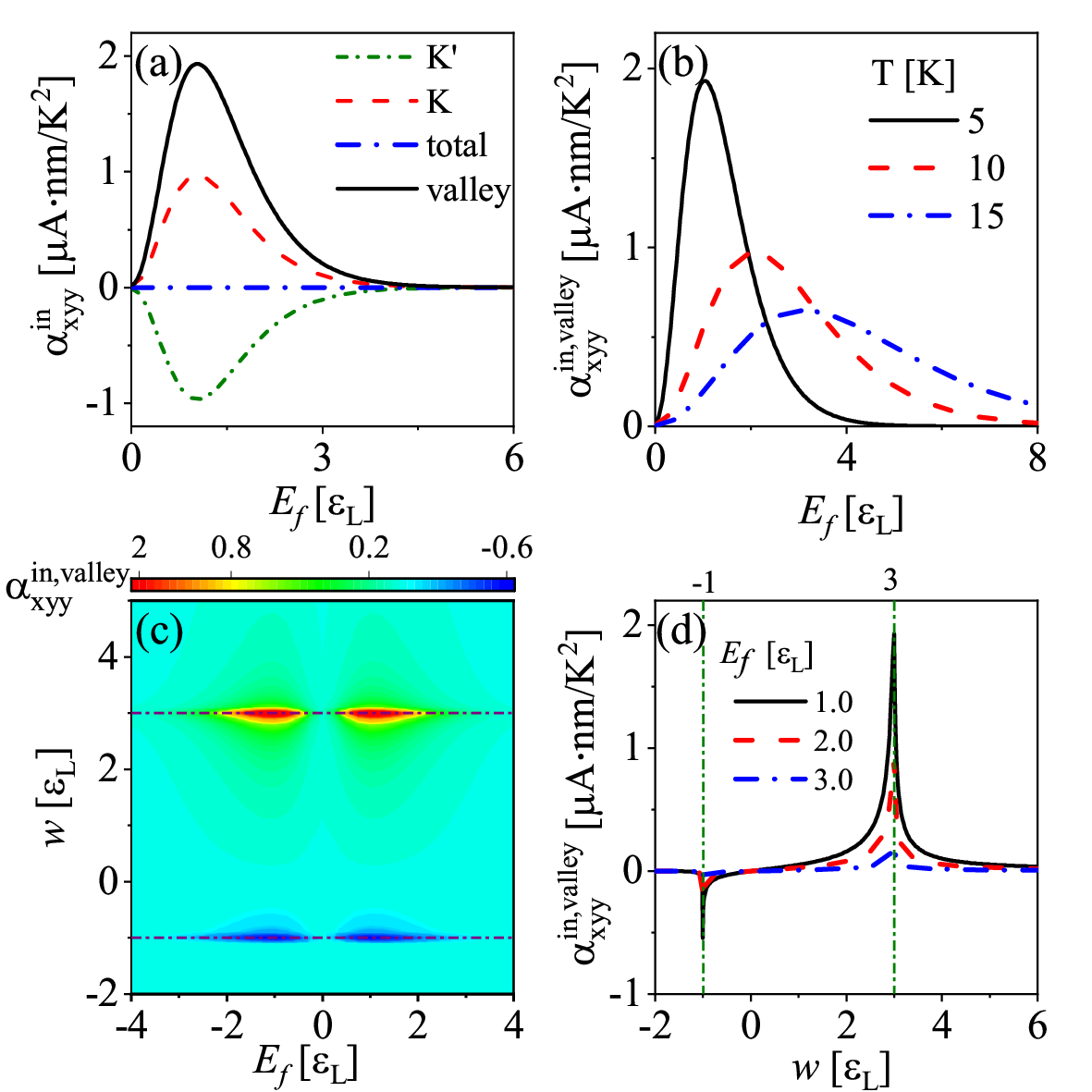}
\caption{ (a) The quantity $\alpha_{xyy}^{\mathrm{nl}}$ as a function of Fermi energy $E_{f}$ for different valleys. (b)The quantity $\alpha_{xyy}^{\mathrm{nl,valley}}$ versus the Fermi energy $E_{f}$ for different temperature $T$. (c) The quantity $\alpha_{xyy}^{\mathrm{nl,valley}}$ as a function of Fermi energy $E_{f}$ and strain parameter $w$. (d) The quantity $\alpha_{xyy}^{\mathrm{nl,valley}}$ versus $w$ for different  $E_{f}$.  Momenta are measured in units of $k_{L}=m^{*}v_{3}/\hbar\approx 0.035 ~ nm^{-1}$ and the effective mass $m^{*}=0.037m_{e}$\cite{M. Koshino}. $T = 5~\mathrm{K}$ is fixed in (a) and (c). $w = 3~\mathrm{meV}$ is taken in (b).  }
\label{fig3}
\end{figure}

\section{Results and Discussion}\label{ReDi}
The global $\mathcal{T}$ and $\mathcal{P}$ symmetries combining with the only surviving local $M_{y}$ symmetry indicate that $\alpha^{\mathrm{in,valley}}_{xyy}$ can be nonzero but require  $\alpha^{\mathrm{in,valley}}_{yxx}=0$ in the uniaxially strained bilayer graphene with the strain along $x$ direction [Table~\ref{table1}].
Therefore, we will only consider the situation in which the temperature gradient is applied along $y$-direction in the following. Additionally, the simultaneous existence of the global $\mathcal{T}$ and $\mathcal{P}$ symmetries will force the ordinary OMMs to vanish and also guarantee the values of temperature-gradient-induced OMMs $m_{z}^{\partial_{y}T}(\boldsymbol{k})$  has opposite sign on the inversion-symmetric side of Fermi energy for different valleys, i.e. $m_{z}^{\partial_{y}T,\mathrm{K}}(\boldsymbol{k})=-m_{z}^{\partial_{y}T,\mathrm{K^{\prime}}}
(-\boldsymbol{k})$, in the strained trigonal-warping bilayer graphene [Figs.~\ref{fig2}(i) and (j)]. Therefore, when applying a temperature gradient along the y-direction in strained trigonal-warping bilayer graphene, the carriers in the inequivalent valleys will not only flow toward opposite edges along $x$-axis due to the nonzero NVNE but also exhibit opposite polarizations of orbital magnetization (see details in Sec.~\ref{TGIOM}). Consequently, accompanying the NVNE, opposite polarizations of the orbital magnetization are generated along opposite edges perpendicular to the temperature gradient in real space, distinguishing the carriers separated by the intrinsic NVNE in real space.

Figure~\ref{fig3}(a) illustrates the dependence of intrinsic nonlinear Nernst coefficients  $\alpha_{xyy}^{\mathrm{in}}$ on the Fermi energy $E_{f}$ for the different valleys in the strained trigonal-warping bilayer
graphene. As expected, the $\alpha_{xyy}^{\mathrm{in}}$ for the $\mathrm{K}$ valley has the same magnitude but opposite sign to that for the $\mathrm{K}^{\prime}$ valley, indicating that the total intrinsic nonlinear Nernst current ($\propto \alpha_{xyy}^{\mathrm{in,K}}+\alpha_{xyy}^{\mathrm{in,K}}$) vanishes but a pure intrinsic nonlinear valley Nernst current ($\propto \alpha_{xyy}^{\mathrm{in,K}}-\alpha_{xyy}^{\mathrm{in,K}}$) can be generated.
In addition, A peak feature of $\alpha_{xyy}^{\mathrm{in,valley}}$ is observed near the Dirac cone which might be ascribed to the fact that the quantum metric elements achieve their maximum in this region [Figs.~\ref{fig2} (c)-(h)]. As the temperature $T$ decreases, the value of the peak increases and its position shifts towards lower energy levels [Fig.~\ref{fig3} (b)].

The behaviour of $\alpha_{xyy}^{\mathrm{nl,valley}}$ are significantly influenced by the strain [Figs.~\ref{fig3} (c) and (d)]. A transition from compressive strain ($w<0$) to tensile strain ($w>0$) results in the sign reversal of $\alpha_{xyy}^{\mathrm{nl,valley}}$, which provides a potential approach for the strain type detection.
Additionally, $\alpha_{xyy}^{\mathrm{nl,valley}}$  exhibits a negative peak at $w=-1~\mathrm{meV}$  followed by a positive peak at $w=3~\mathrm{meV}$)[Figs.~\ref{fig3}(c) and (d)]. 
The appearance of these two peaks might be qualitatively attributed to the effect of strain on the rich Fermi properties of trigonal warping bilayer graphene. For the strain-free ($w=0$) bilayer graphene, a Lifshitz transition occurs at energy $\varepsilon _{L}(\boldsymbol{k})=m^{*}v_{3}^{2}/2\approx 1~ meV$, where the Fermi surface evolves from a single connected pocket into four distinct ones as energy and momentum decrease [Figs. \ref{fig2} (a) and  (b)]: the electronic dispersion consists of one central Dirac cone located at the $\mathrm{K}$ ($\mathrm{K}^{\prime}$) point of the Brillouin zone and three leg Dirac cones. Previous works have demonstrated that the topology of the Fermi surface changes through the merging of Dirac cones when the strain strength is modulated \cite{Y.-L. Wu,R. Battilomo}: for $-\varepsilon_{L}<w\leq 3\varepsilon_{L}$, the electronic dispersion retains four Dirac cones in the Brillounin zone. At $w=-1~\mathrm{meV}$, the two Dirac cones along $k_{x}$ axis merge into single point.  For $w>3~\mathrm{meV}$, in contrast, only the cones along the $k_{x}$ axis are survived, while the other two Dirac cones disappear. Consequently, the observed two peaks precisely correspond to the strain strengths around which the number of Dirac cones decreases.

\section{Conclusion} \label{conclusion}
We study the NVNE  as a second-order response to the temperature gradient. Although the linear valley effect disappears in the systems with both $\mathcal{P}$ and $\mathcal{T}$ symmetries, a intrinsic NVNE, stemming from the quantum metric and independent of the relaxation time, can emerge. Through analyzing the symmetry constraints on this intrinsic NVNE, we identify the largest local symmetry near the valleys in 2D systems for the nonzero intrinsic NVNE is a single local mirror symmetry.
 Additionally, an intrinsic valley-contrasting orbital magnetization, which originates from temperature-gradient-induced OMM, accompanies the intrinsic NVNE and can distinguish the carriers separated by the intrinsic NVNE in real space. The behaviour of the intrinsic NVNE in the uniaxially strained trigonal warping bilayer graphene has been investigated. Our study shows that the intrinsic NVNE can emerge when applying the temperature gradient vertically to the direction of strain. Interestingly, the peaks of the intrinsic NVNE versus strain parameter $w$ precisely correspond to the strain strengths at which the number of Dirac cones decreases. In addition, when transiting the compressive strain $w<0$ into tensile one $w>0$, the sign of intrinsic NVNE changes.

\section{acknowledgements}
This work is supported by the National Natural Science Foundation of China (Grant Nos. 12574073 and 12004107), the National Science Foundation of Hunan, China (Grant No. 2023JJ30118), and the Fundamental Research Funds for the Central Universities.

\bigskip

\textit{Note added}. During the review  process of our manuscript, a complementary and independent work addressing the intrinsic NVNE in a tilted Dirac model and bilayer $\mathrm{WTe_{2}}$\cite{X.-J. Zhang} appeared, which supports our result in the overlapping part.

\appendix

\setcounter{equation}{0}
\setcounter{figure}{0}
\setcounter{table}{0}
\makeatletter
\renewcommand{\theequation}{A\arabic{equation}}
\renewcommand{\thefigure}{A\arabic{figure}}
\renewcommand{\thetable}{A\arabic{table}}

\bigskip
\bigskip

\noindent
\section{The expressions of the nonlinear thermoelectric currents}
\label{Ap-A}
In this section, the expression of nonlinear thermoelectric current density $\boldsymbol{j}^{\mathrm{nl}}$ (where the superscript ``{nl}" represents nonlinear)  as a second-order response to temperature gradient will be derived through the Boltzmann equation and semiclassical dynamical theory up to second order.  Within the relaxation time approximation, the Boltzmann equation in presence of the temperature gradient $\nabla T$ and in absence of the electric field ($\boldsymbol{E}=0$) and magnetic field ($\boldsymbol{B}=0$) can be expressed as

\begin{equation}
\frac{f_{\boldsymbol{k}}-f_{0}}{\tau}=-\frac{\partial f_{\boldsymbol{k}}}{\partial r_{a}}v_{a},
\label{Ap-A-1}
\end{equation}
where $f_{0}=\left(1+e^{(\varepsilon _{\boldsymbol{k}}-E_{f})/(k_{B}T)} \right)^{-1}$ is the equilibrium Fermi distribution, $\tau$ represents relaxation time, $v_{a}$ and $r_{a}$ representing the velocity and coordinate position of electrons in the $a$ direction, respectively. Since we are interested in the response up to the second order in the temperature gradient,
we have the nonequilibrium Fermi distribution function as
\begin{equation}
\begin{aligned}
f(\boldsymbol{r},\boldsymbol{k})\approx f_{0}(\boldsymbol{r},\boldsymbol{k})+\delta f_{1}(\partial_{a}T)+\delta f_{2}(\partial_{a}T\partial_{b}T)
\end{aligned}
\label{Ap-A-2}
\end{equation}
with the terms $\delta f_{n}$ understood to be vanish as ($\partial T/\partial r_{a}$), where $a,b=x$ or $y$, and the local equilibrium Fermi distribution $f_{0}$ depends on $\boldsymbol{r}$ and $\boldsymbol{k}$ via indirect variables $T$ and $\varepsilon _{\boldsymbol{k}}$ as  $f(\boldsymbol{r},\boldsymbol{k})=f_{0}(\varepsilon _{\boldsymbol{k}},T(\boldsymbol{r}))$. Hence, one can  have
\begin{equation}
\begin{aligned}
\frac{\partial f_{0}}{\partial r_{a}}=\frac{\partial f_{0}}{\partial T}\frac{\partial T}{\partial r_{a}}
 =-\frac{(\varepsilon_{\boldsymbol{k}}-E_{f})}{T}\frac{\partial f_{0}}{\partial\varepsilon _{\boldsymbol{k}}}\partial_{a}T.
\end{aligned}
\label{Ap-A-3}
\end{equation}
Accompanying Eq.~(\ref{Ap-A-3}) with the relation  $\partial\varepsilon _{\boldsymbol{k}}/\partial\boldsymbol{k}=\hbar\boldsymbol{v}$, we can express $\partial f_{0}/\partial T$ in terms of $\partial f_{0}/\partial\boldsymbol{k}$ as
\begin{equation}
\begin{aligned}
\frac{\partial f_{0}}{\partial\boldsymbol{k}}=\frac{\partial f_{0}}{\partial\varepsilon _{\boldsymbol{k}}}\frac{\partial\varepsilon _{\boldsymbol{k}}}{\partial\boldsymbol{k}}
 =-\frac{\hbar\boldsymbol{v}T}{(\varepsilon _{\boldsymbol{k}}-E_{f})}\frac{\partial f_{0}}{\partial T}.
\end{aligned}
\label{Ap-A-4}
\end{equation}
Taking the expanded form of $f$ in Eq.~(\ref{Ap-A-2}) into Eq.~(\ref{Ap-A-1}) and comparing the order of temperature gradient for both sides of the equation, one has
\begin{equation}
\begin{aligned}
 &\delta f_{1}=-\tau\frac{\partial f_{0}}{\partial T}v_{a}\partial_{a}T,\\
 &\delta f_{2}=\tau^{2}\left(\frac{\partial^{2}f_{0}}{\partial T^{2}}\partial_{a}T\partial_{b}T+\frac{\partial f_{0}}{\partial T}\partial_{ab}T\right)v_{a}v_{b},
\end{aligned}
\end{equation}
where we have introduce the shorthand notations $\partial_{a}=\partial/\partial r_{a}$ and $\partial_{ab}=\partial/\partial r_{a}\partial r_{b}$ for the simplification. When assuming the temperature gradient is uniform (i.e. $\partial_{ab}T=0$), the first-order $\delta f_{1}$  (second-order $\delta f_{2}$) correction of the temperature gradient to the non-equilibrium distribution functions is found to be
\begin{equation}
\begin{aligned}
 &\delta f_{1} =\frac{\tau}{T\hbar}(\varepsilon _{\boldsymbol{k}}-E_{f})\frac{\partial f_{0}}{\partial k_{a}}\partial_{a}T,\\
 &\delta f_{2} =\tau^{2}\left[2\hbar v_{b}\frac{\partial f_{0}}{\partial k_{a}}+(\varepsilon _{\boldsymbol{k}}-E_{f})\frac{\partial^{2}f_{0}}{\partial k_{a}\partial k_{b}} \right]  \\
   &~~~~~~~~~~~ \times\frac{\varepsilon _{\boldsymbol{k}}-E_{f}}{\hbar^{2}T^{2}}\partial_{a}T\partial_{b}T.
\label{fk}
\end{aligned}
\end{equation}
Besides, when further applying the temperature gradient in a single direction, one can have $a=b$.

Gao \textit{el at.} have derived the expression of thermoelectric current density $\boldsymbol{j}$ from a single Bloch band labeled by $0$ through the semiclassical framework of electron dynamics up to second order. \cite{Y. Gao1} One can easily expand their formula to the multiple-band case, in which multiple-Bloch band contribution to thermoelectric current density is considered, by changing the label $0$ into $n$ (band index) and then taking a summation over $n$. Therefore, the total thermoelectric current density $\boldsymbol{j}$, taking the conventional term $-e\sum_{n}\int[d\boldsymbol{k}]\boldsymbol{v}^{n}f_{\boldsymbol{k}}$ from the carrier's group velocity into consideration, is found to be
\begin{equation}
\begin{aligned}
\boldsymbol{j}=  &-e\sum_{n}\int[d\boldsymbol{k}]\boldsymbol{v}^{n}f_{\boldsymbol{k}}
 -\frac{\nabla T}{T}\times\frac{e}{\hbar}\sum_{n}\int[d\boldsymbol{k}]{\boldsymbol{\Omega}^{n}}
 (\boldsymbol{k})\\
  &\times \left((\varepsilon ^{n}_{\boldsymbol{k}}-E_{f})f_{\boldsymbol{k}}+\frac{1}{k_{B}T}
  \mathrm{ln}(1+e^{\frac{\varepsilon ^{n}_{\boldsymbol{k}}-E_{f}}{k_{B}T}}) \right)  \\
  &+\nabla T\times\frac{e}{\hbar}\sum_{n}\int [d\boldsymbol{k}]
  \frac{(\varepsilon ^{n}_{\boldsymbol{k}}-E_{f})}{T^{2}}\frac{\partial
  f_{\boldsymbol{k}}}{\partial \varepsilon ^{n}_{\boldsymbol{k}}}\\
   & \times\left(2\hat{e}_{a}\hbar\sum_{m\neq n} \mathrm{Re}\frac{\mathcal{A}_{a}^{nm}(\boldsymbol{v}^{n}\times\boldsymbol{\mathcal{A}}^{mn})_{b}}
   {\varepsilon _{\boldsymbol{k}}^{n}-\varepsilon _{\boldsymbol{k}}^{m}}\partial_{b}T \right),\\
\label{j-1}
\end{aligned}
\end{equation}
where $\boldsymbol{\Omega}^{n}(\boldsymbol{k})=\nabla_{\boldsymbol{k}}\times\boldsymbol{\mathcal{A}}^{n}
(\boldsymbol{k})$ is the  unperturbed BC, $\boldsymbol{\mathcal{A}}^{n}(\boldsymbol{k})=\langle u_{n}(\boldsymbol{k})|i\nabla_{\boldsymbol{k}}|u_{n}(\boldsymbol{k})\rangle$ denotes the intraband Berry connection with $|u_{n}(\boldsymbol{k})\rangle$ being the periodic part of the Bloch wave-function, $\boldsymbol{\mathcal{A}}^{nm}(\boldsymbol{k})=\langle u_{n}(\boldsymbol{k})|i\nabla_{\boldsymbol{k}}|u_{m}(\boldsymbol{k})\rangle$ representing the interband Berry connection, $\boldsymbol{v}^{mn}(\boldsymbol{k})=\langle u_{m}(\boldsymbol{k})|\hat{\boldsymbol{v}}|u_{n}(\boldsymbol{k})\rangle$ and $\boldsymbol{v}^{n}(\boldsymbol{k})=\langle u_{n}(\boldsymbol{k} |\hat{\boldsymbol{v}}|u_{n}(\boldsymbol{k})\rangle$
 indicating the interband and intraband velocity matrix element, respectively.
The second term is ascribed to the BC
$\boldsymbol{\Omega}^{n}(\boldsymbol{k})$, and the third term stems from the orbital magnetic quadrupole moment density. Obviously, both second and third terms have the quantum origins from the geometric properties of the electron wave function.

Taking the formula of the nonequilibrium distribution function $f_{\boldsymbol{k}}$ determined by
Eqs.(\ref{Ap-A-2}) and (\ref{fk}), the formula of the nonlinear thermoelectric current density $j^\mathrm{nl}_{a}$ along $a$ direction as a second-order response to the temperature gradient in the 2D systems is determined as
\begin{widetext}
\begin{equation}
\begin{aligned}
j^{\mathrm{nl}}_{a}=&\left\{-\frac{e\tau^{2}}{\hbar^{2}}\sum_{n}\int[d\boldsymbol{k}]v_{a}^{n}
\frac{\left(\varepsilon ^{n}_{\boldsymbol{k}}-E_{f}\right)}{T^{2}}
\left[2\hbar v_{c}^{n}\frac{\partial f_{0}}{\partial k_{b}}
+\left(\varepsilon ^{n}_{\boldsymbol{k}}-E_{f}\right)\frac{\partial^{2}f_{0}}{\partial k_{b}\partial k_{c}} \right]-\frac{e\tau}{\hbar} \epsilon_{abd}\sum_{n}\int[d\boldsymbol{k}]\frac{\left(\varepsilon ^{n}_{\boldsymbol{k}}-E_{f}\right)^{2}}{T^{2}}
 v_{c}^{n}\frac{\partial f_{0}}{\partial \varepsilon ^{n}_{\boldsymbol{k}}}\Omega_{d}^{n}(\boldsymbol{k})\right.\\
  &\left.+e\sum_{m\neq n}^{n}\int[d\boldsymbol{k}]\frac{\left(\varepsilon ^{n}_{\boldsymbol{k}}-E_{f}\right)^{2}}{T^{2}}\frac{\partial f_{0}}{\partial \varepsilon ^{n}_{\boldsymbol{k}}}
 \left[\frac{2v_{a}^{n}\mathcal{G}_{bc}^{nm}-v_{b}^{n}\mathcal{G}_{ac}^{nm}-
 v_{c}^{n}\mathcal{G}_{ab}^{nm}}{\varepsilon _{\boldsymbol{k}}^{n}-
 \varepsilon _{\boldsymbol{k}}^{m}}\right]\right\}\partial_{b}T\partial_{c}T,
\end{aligned}
\label{App-A-TO}
\end{equation}
\end{widetext}
where $\mathcal{G}_{ad}^{nm}(\boldsymbol{k})=\mathrm{Re}[\mathcal{A}_{a}^{nm}\mathcal{A}_{d}^{mn}]$ is  the band-resolved quantum metric\cite{Wang1,Zhuang,N.-Z. Wang,A. Gao}, which quantifies the distance between wave functions in the parameter space.
 The nonlinear thermoelectric currents ${j}^\text{nl}_{a}$ in $a$ direction can be further decomposed into three components as
 ${j}^{\mathrm{nl}}_{a} ={j}^{\mathrm{nl}}_{1,a}(\tau^{0})+{j}^{\mathrm{nl}}_{2,a}(\tau^{1})+
{j}^{\mathrm{nl}}_{3,a}(\tau^{2})$ based on the dependence on the scale of the relaxation time $\tau$. Therefore, the corresponding nonlinear thermoelectric coefficients $\alpha_{abc,i(=1,2,3)}$, which are defined as $j^{\mathrm{nl}}_{i,a}=\alpha_{abc,i}\partial_{b}T\partial_{c}T$, are found to be, respectively,
\begin{align}
\alpha_{abc,1}&=e \sum^{n}_{m\neq n}\int[d\boldsymbol{k}]\frac{(\varepsilon _{\boldsymbol{k}}^{n}-E_{f})^{2}}{T^{2}}\frac{\partial f_{0}}{\partial \varepsilon _{\boldsymbol{k}}^{n}}\nonumber \\
  &\times\left[\frac{2v_{a}^{n}\mathcal{G}_{bc}^{nm}-v_{b}^{n}\mathcal
  {G}_{ac}^{nm}-v_{c}^{n}\mathcal{G}_{ab}^{nm}}{\varepsilon _{\boldsymbol{k}}^{n}-\varepsilon _{\boldsymbol{k}}^{m}}\right],
\label{Ap-A-tau-0}\\
\alpha_{abc,2}\!\!&=-\frac{e\tau}{\hbar}\epsilon_{abd}  \sum_{n}\!\!\int\!\![d\boldsymbol{k}]
 \frac{(\!\varepsilon ^{n}_{\boldsymbol{k}}\!-\!E_{f}\!)^{2}}{T^{2}}
v_{c}^{n}\frac{\partial f_{0}}{\partial \varepsilon ^{n}_{\boldsymbol{k}}}\Omega^{n}_{d}\!(\boldsymbol{k}\!),
\label{Ap-A-tau-1}\\
\alpha_{abc,3}&=\frac{e\tau^{2}}{T^{2}}\sum_{n}\int[d\boldsymbol{k}]\frac{(\varepsilon ^{n}_
{\boldsymbol{k}}-E_{f})^{2}}{m_{ab}(\boldsymbol{k})}v_{c}^{n}\frac{\partial f_{0}}{\partial \varepsilon ^{n}_{\boldsymbol{k}}}
\label{Ap-A-tau-1}
\end{align}
with $m_{ab}^{-1}(\boldsymbol{k})=(1/\hbar)(\partial v_{a}/\partial k_{b})$ indicating the inverse effective mass tensor. Obviously, the nonlinear thermoelectric coefficients  $\alpha_{abc,1}\propto \tau^{0}$, $\alpha_{abc,2} \propto \tau$ and $\alpha_{abc,3} \propto \tau^{2}$ are disclosed to originate from the quantum metric, BC, and Drude contributions, respectively. Hence, the coefficients [$\alpha_{abc,1}$, $\alpha_{abc,2}$, $\alpha_{abc,3}$] have been rewritten in the main text as [$\alpha^{\text{in}}_{abc}$, $\alpha^{\text{B}}_{abc}$, $\alpha^{\text{D}}_{abc}$], where superscript ``{in}",  ``{B}", and  ``{D}"  represents intrinsic, Berry, and Drude, respectively.

\makeatletter
\renewcommand{\theequation}{B\arabic{equation}}
\renewcommand{\thefigure}{B\arabic{figure}}
\renewcommand{\thetable}{B\arabic{table}}
\section{The exploitation of the symmetry constraints }
\label{Ap-B}
The symmetry constraints on the coefficients [$\varepsilon ^{n}(\boldsymbol{k})$ $\boldsymbol{\mathcal{A}}^{nm}(\boldsymbol{k})$, $\boldsymbol{v}^{n}\left(\boldsymbol{k}\right)$, $\mathcal{G}_{ab}^{mn}(\boldsymbol{k})]$ from $\mathcal{T}$ / $\mathcal{P}$ symmetries  will be exploited in the section. We firstly analyse the $\mathcal{T}$-symmetric constraint.
In spin-less $\mathcal{T}$-symmetric systems, the time-reversal operator is expressed as $\mathcal{T}=\mathcal{K}$ determined by the complex conjugation operator $\mathcal{K}$, and the effective Hamiltonian $H(\boldsymbol{k})$ and the periodic part of the Bloch wave-function $|u_{n}(\boldsymbol{k})\rangle$ satisfy
\begin{equation}
\begin{aligned}
&\mathcal{T} H(\boldsymbol{k})\mathcal{T}^{-1}=H(-\boldsymbol{k}), \\
&|u_{n}(-\boldsymbol{k})\rangle=|u_{n}(\boldsymbol{k})\rangle^{*},
\end{aligned}
\end{equation}
which gives the energy eigenvalue $\varepsilon ^{n}(\boldsymbol{k})$ of the effective Hamiltonian ($H(\boldsymbol{k})|u_{n}(\boldsymbol{k})\rangle=\varepsilon ^{n}(\boldsymbol{k})|u_{n}
(\boldsymbol{k})\rangle$) is an even function of the momentum $\boldsymbol{k}$, namely
\begin{equation}
\varepsilon ^{n}(\boldsymbol{k})=\varepsilon ^{n}(-\boldsymbol{k}),
\end{equation}
enforcing the group velocity $\boldsymbol{v}^{n}(\boldsymbol{k})=(1/\hbar)(\partial\varepsilon ^{n}/\partial\boldsymbol{k})$ odd and the inverse effective mass tensor $m_{ab}^{-1}(\boldsymbol{k})=(1/\hbar)(\partial v_{a}^{n}/\partial k_{b})$ even with respective to $\boldsymbol{k}$. The interband Berry connection satisfies:
\begin{equation}
\begin{aligned}
\boldsymbol{\mathcal{A}}^{mn}(\boldsymbol{-k})&=\langle u_{m}(-\boldsymbol{k})|i\partial_{-\boldsymbol{k}}|u_{n}(-\boldsymbol{k})\rangle \\
&=-i\langle u_{m}(\boldsymbol{k})^{*}|\partial_{\boldsymbol{k}}u_{n}(\boldsymbol{k})^{*}
\rangle\\
&=-i\langle\partial_{\boldsymbol{k}}u_{n}(\boldsymbol{k})|u_{m}(\boldsymbol{k})
\rangle\\
&=i\langle u_{n}(\boldsymbol{k})|\partial_{\boldsymbol{k}}u_{m}(\boldsymbol{k})
\rangle\\
&=\boldsymbol{\mathcal{A}}^{nm}(\boldsymbol{k}).
\end{aligned}
\end{equation}
To obtaining the last second equality, we have used the relation $\langle \partial_{\boldsymbol{k}} u_{n}(\boldsymbol{k})|u_{m}(\boldsymbol{k})\rangle+\langle u_{n}(\boldsymbol{k})|\partial_{\boldsymbol{k}} u_{m}(\boldsymbol{k})\rangle=\partial_{\boldsymbol{k}}\langle u_{n}(\boldsymbol{k})|u_{m}(\boldsymbol{k})\rangle=0$.
Consequently, when the systems is $\mathcal{T}$-symmetric, BC $\Omega^{nm}_{ab}(\boldsymbol{k})$ and the band-resolved quantum metric $\mathcal{G}^{nm}_{ab}(\boldsymbol{k})$ satisfy
\begin{equation}
\begin{aligned}
\Omega^{nm}_{ab}(\boldsymbol{k})&=i[\mathcal{A}^{nm}_{a}(\boldsymbol{k})\mathcal{A}^{mn}_{b}
(\boldsymbol{k})-\mathcal{A}^{nm}_{b}(\boldsymbol{k})\mathcal{A}^{mn}_{a}(\boldsymbol{k})]\\
&=i[\mathcal{A}^{mn}_{a}(-\boldsymbol{k})\mathcal{A}^{nm}_{b}
(\boldsymbol{-k})-\mathcal{A}^{mn}_{b}(\boldsymbol{k})\mathcal{A}^{nm}_{a}(-\boldsymbol{k})]\\
&=i[\mathcal{A}^{nm}_{b}(\boldsymbol{-k})\mathcal{A}^{mn}_{a}(-\boldsymbol{k})-
\mathcal{A}^{nm}_{a}(-\boldsymbol{k})\mathcal{A}^{mn}_{b}(\boldsymbol{k})]\\
&=-\Omega^{nm}_{ab}(-\boldsymbol{k})
\end{aligned}
\label{app-A-C1}
\end{equation}
and
\begin{equation}
\begin{aligned}
\mathcal{G}^{nm}_{ab}(\boldsymbol{k}) &=\frac{1}{2} \mathrm{Re}[\mathcal{A}_{a}^{nm}(\boldsymbol{k})\mathcal{A}_{b}^{mn}(\boldsymbol{k})
+\mathcal{A}_{b}^{nm}(\boldsymbol{k})\mathcal{A}_{a}^{mn}(\boldsymbol{k})]\\
&=\frac{1}{2} \mathrm{Re}[\mathcal{A}_{a}^{mn}(-\boldsymbol{k})\mathcal{A}_{b}^{nm}(-\boldsymbol{k})
+\mathcal{A}_{b}^{mn}(-\boldsymbol{k})\mathcal{A}_{a}^{nm}(-\boldsymbol{k})]\\
&=\frac{1}{2} \mathrm{Re}[\mathcal{A}_{b}^{nm}(-\boldsymbol{k})\mathcal{A}_{a}^{mn}(-\boldsymbol{k})
+\mathcal{A}_{a}^{nm}(-\boldsymbol{k})\mathcal{A}_{b}^{mn}(-\boldsymbol{k})]\\
&=\mathcal{G}^{nm}_{ab}(-\boldsymbol{k}).
\end{aligned}
\end{equation}

Next, let's turn our sight into the symmetry constraint from $\mathcal{P}$ symmetry.  Under $\mathcal{P}$ operation, the position vector $\boldsymbol{r}$ changes sign to $-\boldsymbol{r}$,
and the effective Hamiltonian $H(\boldsymbol{k})$ and the corresponding periodic part of the Bloch wave-function $|u_{n}(\boldsymbol{k})\rangle$ satisfy:
\begin{equation}
\begin{aligned}
&\mathcal{P} H(\boldsymbol{k})\mathcal{P}^{-1}=H(-\boldsymbol{k}), \\
&\mathcal{P}|u_{n}(\boldsymbol{k})\rangle=|u_{n}(-\boldsymbol{k})\rangle,
\end{aligned}
\end{equation}
the energy eigenvalue of the Bloch Hamiltonian has following symmetry with respect to $\boldsymbol{k}$
\begin{equation}
\varepsilon ^{n}(\boldsymbol{k})=\varepsilon ^{n}(-\boldsymbol{k}),
\end{equation}
the symmetries of group velocity $\boldsymbol{v}^{n}(\boldsymbol{k})$ and the inverse effective mass tensor $m_{ab}^{-1}(\boldsymbol{k})$ are same to those under $\mathcal{T}$ symmetry, namely $\boldsymbol{v}^{n}(\boldsymbol{k})\xrightarrow{\mathcal{P}}-\boldsymbol{v}^{n}
(-\boldsymbol{k})$ and $m_{ab}^{-1}(\boldsymbol{k})\xrightarrow{\mathcal{P}}m_{ab}^{-1}(\boldsymbol{-k})$.
The interband Berry connection $\boldsymbol{\mathcal{A}}^{nm}(\boldsymbol{k})$ in the  $\mathcal{P}$-symmetric systems meets

\begin{equation}
\begin{aligned}
\boldsymbol{\mathcal{A}}^{nm}(\boldsymbol{k})&=\langle u_{n}(\boldsymbol{k})|i\partial_{\boldsymbol{k}}|u_{m}(\boldsymbol{k})\rangle \\
&=i\langle u_{n}(\boldsymbol{k})|\mathcal{P}^{+}\mathcal{P}\partial_{\boldsymbol{k}}|u_{m}(\boldsymbol{k})
\rangle \\
&=-i\langle u_{n}(-\boldsymbol{k})|\partial_{\boldsymbol{k}}|u_{m}(-\boldsymbol{k})\rangle\\
 &=-\boldsymbol{\mathcal{A}}^{nm}(-\boldsymbol{k}).
\end{aligned}
\end{equation}
Therefore, Berry curvature and the quantum metric follow
\begin{equation}
\begin{aligned}
\Omega^{nm}_{ab}(\boldsymbol{k})&=i[\mathcal{A}^{nm}_{a}(\boldsymbol{k})\mathcal{A}^{mn}_{b}
(\boldsymbol{k})-\mathcal{A}^{nm}_{b}(\boldsymbol{k})\mathcal{A}^{mn}_{a}(\boldsymbol{k})]  \\
&=i[\mathcal{A}^{nm}_{a}(-\boldsymbol{k})\mathcal{A}^{mn}_{b}
(-\boldsymbol{k})-\mathcal{A}^{nm}_{b}(-\boldsymbol{k})\mathcal{A}^{mn}_{a}(-\boldsymbol{k})]\\
&  =\Omega^{nm}_{ab}(-\boldsymbol{k})
\end{aligned}
\label{app-A-C2}
\end{equation}
and
\begin{equation}
\begin{aligned}
\mathcal{G}_{ab}^{nm}(\boldsymbol{k}) &= \frac{1}{2}\mathrm{Re}[\mathcal{A}_{a}^{nm}(\boldsymbol{k})\mathcal{A}_{b}^{mn}(\boldsymbol{k})
+\mathcal{A}_{b}^{nm}(\boldsymbol{k})\mathcal{A}_{a}^{mn}(\boldsymbol{k})] \\
&=\frac{1}{2}\mathrm{Re}[\mathcal{A}_{a}^{nm}(-\boldsymbol{k})\mathcal{A}_{b}^{mn}(-\boldsymbol{k})
+\mathcal{A}_{b}^{nm}(-{k})\mathcal{A}_{a}^{mn}(-\boldsymbol{k})] \\
& = \mathcal{G}_{ab}^{nm}(-\boldsymbol{k}).
\end{aligned}
\end{equation}
According to Eqs.~(\ref{app-A-C1}) and (\ref{app-A-C2}), one can find that $\Omega^{nm}_{ab}(\boldsymbol{k})=0$ when in presence of both $\mathcal{T}$ and $\mathcal{P}$ symmetries.

The $a$-th component of the temperature-gradient induced OMM $m_{n,a}^{\nabla T}(\boldsymbol{k})$  for the $n-$th band is written as
\begin{equation}
\begin{aligned}
m_{n,a}^{\nabla T}(\boldsymbol{k})
& =\sum_{m\neq n }\left[2\mathrm{Re}\frac{\mathcal{M}_{a}^{nm}\mathcal{A}_{d}^{mn}}{\varepsilon ^{n}_{\boldsymbol{k}}
-\varepsilon ^{m}_{\boldsymbol{k}}}
+\frac{e}{2\hbar}\epsilon_{abc}(\partial_{b}\mathcal{G}_{cd}^{mn}) \right]\partial_{d} T   \\
&= \epsilon_{abc}\left[e\sum_{l\neq m}\sum_{m\neq n}\frac{\mathrm{Re}[ v_{b}^{nl}A_{c}^{lm}A_{d}^{mn}]}{\varepsilon ^{n}_{\boldsymbol{k}}-\varepsilon ^{m}_{\boldsymbol{k}}}  \right.  \\
& \left.+e\sum_{m\neq n}\frac{\mathrm{Re}[v_{b}^{m}A_{c}^{lm}A_{d}^{mn}]}{\varepsilon ^{n}_{\boldsymbol{k}}-
     \varepsilon ^{m}_{\boldsymbol{k}}}
                                +\frac{e}{2\hbar}\epsilon_{abc}(\partial_{b}\mathcal{G}_{cd}^{mn}) \right]\partial_{d} T ,  \\
\end{aligned}
\end{equation}
Based on the symmetry constraints of the coefficients [$\varepsilon ^{n}(\boldsymbol{k})$, $\boldsymbol{v}^{n}(\boldsymbol{k})$, $\boldsymbol{\mathcal{A}}^{nm}(\boldsymbol{k})$, $\mathcal{G}^{mn}_{ab}(\boldsymbol{k})$ and $\partial_{d}T$], the temperature-gradient induced OMM $m_{n,a}^{\nabla T}(\boldsymbol{k})$ are, accordingly, transformed as
\begin{equation}
m_{n,a}^{\nabla T}(\boldsymbol{k})=-m_{n,a}^{\nabla T}(-\boldsymbol{k})
\end{equation}
for both $\mathcal{T}$ and $\mathcal{P}$ symmetries.
The above analysed transformations of the coefficients under $\mathcal{P}$/$\mathcal{T}$ symmetries are summarized in Table \ref{App-B-table}.

\begin{table}[htp]
\centering
\caption{The constraints on the different quantities under  time-reversal ($\mathcal{T}$) and inversion ($\mathcal{P}$) symmetries.}
\vspace{5pt}
\setlength{\tabcolsep}{4mm}{
\begin{tabular}{ccc}
\toprule
                                                &              $\mathcal{T} $                 &              $\mathcal{P} $                                   \\
\hline
       $\partial_{d}T$        &    $-\partial_{d}T$       &    $\partial_{d}T$                                                          \\
       $\varepsilon ^{n}(\boldsymbol{k})$        &    $\varepsilon ^{n}(-\boldsymbol{k})$       &    $\varepsilon ^{n}(-\boldsymbol{k})$                                  \\
       $\boldsymbol{v}^{n}(\boldsymbol{k})$     &   $-\boldsymbol{v}^{n}(-\boldsymbol{k})$    &    $-\boldsymbol{v}^{n}(-\boldsymbol{k})$                                          \\
       $m_{ab}^{-1}(\boldsymbol{k})$             &   $m_{ab}^{-1}(-\boldsymbol{k})$              &   $m_{ab}^{-1}(-\boldsymbol{k})$                                                  \\
   $\boldsymbol{\mathcal{A}}^{nm}(\boldsymbol{k})$&$\boldsymbol{\mathcal{A}}^{mn}(-\boldsymbol{k})$&$-\boldsymbol{\mathcal{A}}^{nm}(-\boldsymbol{k})$  \\
       $\Omega^{nm}_{ab}(\boldsymbol{k})$       &  $-\Omega^{nm}_{ab}(-\boldsymbol{k})$       &    $\Omega^{nm}_{ab}(-\boldsymbol{k})$                     \\
       $\mathcal{G}^{nm}_{ab}(\boldsymbol{k})$ &  $\mathcal{G}^{nm}_{ab}(-\boldsymbol{k})$   &    $\mathcal{G}^{nm}_{ab}(-\boldsymbol{k})$         \\
       $\mathbf{m}_{n}^{\nabla T}(\boldsymbol{k})$   &  $-\mathbf{m}_{n}^{\nabla T}(-\boldsymbol{k})$   &   $-\mathbf{m}_{n}^{\nabla T}(-\boldsymbol{k})$                     \\
\hline
\end{tabular} }
\label{App-B-table}
\end{table}

\makeatletter
\renewcommand{\theequation}{C\arabic{equation}}
\renewcommand{\thefigure}{C\arabic{figure}}
\renewcommand{\thetable}{C\arabic{table}}
\section{An simplified experimental setup for observing and detecting the intrinsic nonlinear valley Nernst effect in bilayer graphene}
\label{App-C}
In this Appendix, we will propose a simplified experimental setup to achieve the NVNE in the bilayer graphene and discuss an experimental scenario to detect the NVNE in the systems with both $\mathcal{T}$ and $\mathcal{P}$ symmetries.

\begin{figure}[htbp]
\centering
\flushleft
\includegraphics[width=1.0\linewidth,clip]{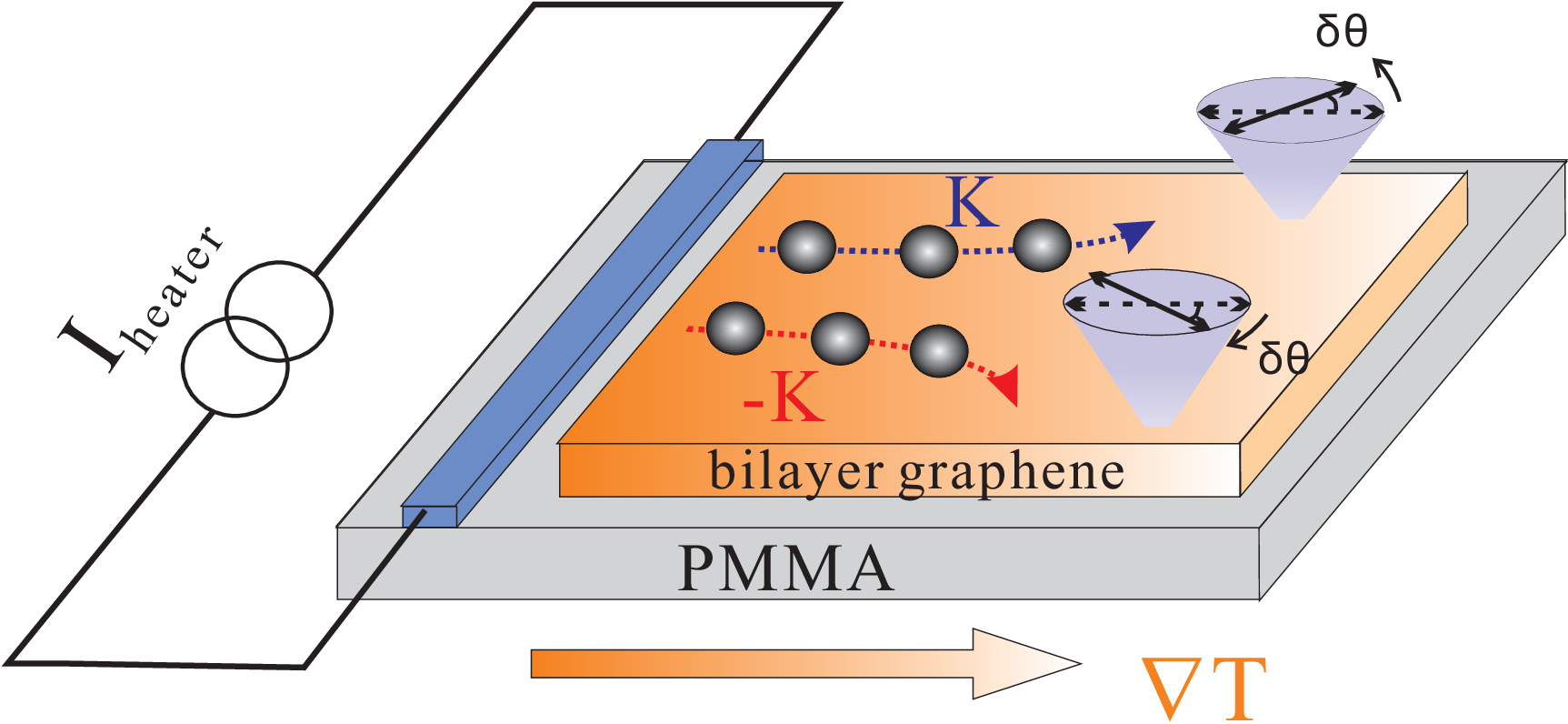}
\caption{Schematics of the proposed device for the nonlinear valley Nernst effect (NVNE) in the strained graphene, showing the bilayer graphene placed on the flexible polymethyl methacrylate (PMMA) substrate and a local heater electrode generates a temperature gradient. NVNE can be detected through the magneto-optical Kerr method, namely focusing a linearly polarized probe beam onto the device under normal incidence and measuring the Kerr rotation angle $\delta\theta$ of the reflected beam. }
\label{figC1}
\end{figure}

To achieve the NVNE ($\propto(\nabla T)^{2}$) in the bilayer graphene, it's critical to ensure the  feasibility of simultaneously applying a high temperature gradient and strain within a simple experimental setup. Based on the well-established techniques in the thermal detection \cite{J. Xu2019,Liu-2025} and strain engineering, we propose a simplified experimental setup (Fig.~\ref{figC1}) to realize NVNE for experiment.

As illustrated in Fig.~\ref{figC1}, the high temperature gradient can be generated by applying a direct current to local heater electrode\cite{J. Xu2019,S. M. Wu2016}. The temperature gradient from local heater electrodes can actually reach 1.5 $\mathrm{K/\mu m}$\cite{J. Xu2019}. It should be mentioned that one can also realize the high temperature gradient by shining a laser beam to heat up one side of the sample\cite{P. Farahmand2015}. Additionally, the uniaxial strain, which is required to break $C_3$ symmetry of bilayer graphene (see Sec. \ref{EHSA}), can be induced through mechanical bending by placing the material on flexible substrates\cite{J. Son2019,M. Huang2009,T. Yu2008} like  PMMA. These elastic polymer substrates permit controllable and reversible strain tuning, allowing the systematic investigations into the strain effect on NVNE.

After proposing the simplified experimental setup to achieve the NVNE, the next task is how to probe the NVNE experimentally. As discussed in the main text (Sec.~\ref{TGIOM}), an extra VCOM, which originates from the temperature-gradient-corrected OMM, will arise in the systems with both $\mathcal{T}$ and $\mathcal{P}$ symmetries, where both conventional orbital magnetization and linear Nernst current are suppressed to zero. Consequently, when applying the temperature gradient $\nabla T$ to $\mathcal{P}$- and $\mathcal{T}$- symmetric materials, the flowing of inequivalent valleys toward opposite edges perpendicular to $\nabla T$ not only gives rise to valley current but also leads to the generation of opposite polarization of the orbital magnetization along opposite edges.  Therefore, the magneto-optical Kerr method (MOKM)\cite{J. Lee2016,J. Son2019}, a technique successfully applied to detect the valley Hall effect\cite{J. Lee2016} and Berry curvature dipole\cite{J. Son2019}, is suitable to probe the NVNE.
In this method, linearly polarized light is incident normally onto the sample, and the  polarization of the reflected beam experiences a rotation angle proportional to the out-of-plane magnetization of the materials. Hence, when applied to our proposed effect, MOKM would not only detect a nonzero Kerr rotation angle $\delta \theta$ of the reflected beam but also reveal opposite signs of $\delta \theta$ at opposing edges of the sample [Fig. \ref{figC1}]. Those observation will directly reflect VCOM induced by the temperature gradient and imply the emergence of valley current.

It should be mentioned that Zhang \textit{et al.} \cite{X.-J. Zhang} recently demonstrated that nonlocal transport measurement can be employed to detect NVNE when the materials exhibit $\mathcal{P}$ symmetry breaking. In such nonlocal measurements, the breaking of $\mathcal{P}$ symmetry of the sample is required, as the involved inverse linear and nonlinear Hall effects vanish in centrosymmetric crystals \cite{X.-J. Zhang,J. Cao2025}.

\end{document}